\documentclass{article}
\usepackage{waspaa23,amsmath,graphicx,url,times}
\usepackage{color}
\usepackage{multirow}
\usepackage{amssymb}

\usepackage{ifthen}
\title{Pretraining respiratory sound representations using metadata and contrastive learning}

\name{Ilyass Moummad, Nicolas Farrugia}%\thanks{This work was co-funded by the AI@IMT program of ANR (French National Agency) and the OSO-AI company.}}
\address{IMT Atlantique, Lab-STICC, UMR CNRS 6285, F-29238 Brest, France}
\begin{document}

\ninept
\maketitle

\begin{sloppy}

\begin{abstract}
Methods based on supervised learning using annotations in an end-to-end fashion have been the state-of-the-art for classification problems. However, they may be limited in their generalization capability, especially in the low data regime. In this study, we address this issue using supervised contrastive learning combined with available metadata to solve multiple pretext tasks that learn a good representation of data. We apply our approach on respiratory sound classification. This task is suited for this setting as demographic information such as sex and age are correlated with presence of lung diseases, and learning a system that implicitly encode this information may better detect anomalies. Supervised contrastive learning is a paradigm that learns similar representations to samples sharing the same class labels and dissimilar representations to samples with different class labels. The feature extractor learned using this paradigm extract useful features from the data, and we show that it outperforms cross-entropy in classifying respiratory anomalies in two different datasets. We also show that learning representations using only metadata, without class labels, obtains similar performance as using cross entropy with those labels only. In addition, when combining class labels with metadata using multiple supervised contrastive learning, an extension of supervised contrastive learning solving an additional task of grouping patients within the same sex and age group, more informative features are learned. This work suggests the potential of using multiple metadata sources in supervised contrastive settings, in particular in settings with class imbalance and few data.
\end{abstract}


\section{Introduction}
\label{sec:intro}

The main idea of Self-Supervised Learning (SSL) is to solve a pretext task to learn a feature extractor producing useful representations without using labels. Contrastive methods are a family of SSL approaches that optimize an encoder to output similar embeddings for different views of the same data, relying on data augmentations (e.g. SimCLR~\cite{chen2020simple}). 
In audio, prior work using contrastive approaches have proposed to use synthetic mixing of training examples to generate views~\cite{fonseca2021unsupervised}, as well as sound separation~\cite{fonseca2021self}, or generating pairs by sampling segments from audio clips~\cite{cola}. 

Constrastive learning can also be successfully exploited using a framework called Supervised Contrastive Learning (SCL), which uses classification labels only to group positive pairs in a contrastive setting similar to SimCLR, resulting in state of the art classification on several computer vision benchmarks~\cite{khosla2020supervised}. SCL has also been applied for environmental sound classification~\cite{soundclr}. Instead of using class labels, SCL could be applied on other available information to define pretext tasks. Practical applications usually include metadata, e.g. demographics in medical applications. In this paper, we explore this idea for respiratory sound classification.

Respiratory sound classification is the task of identifying the diagnosis of a breathing cycle whether it's normal or abnormal. Diagnosing a respiratory pathology using ML on audio data would reduce the work overload for physicians and medical experts, and make medical examination less prone to error. ICBHI~\cite{rocha2019open} and SPRSound~\cite{SPRSound} are public datasets for distinguishing between normal breathing, and respiratory anomalies such as crackle, wheeze, rhonchi and stridor. The two datasets contain recordings of thousands of breathing cycles of varying durations, with an imbalanced class distribution. SPRS recordings were made with an electronic sthetoscope on four different locations, as for ICBHI, recordings were made with four different devices (microphone and electronic stethoscopes) on seven different chest locations. These properties make respiratory sound classification a challenging task. Recent methods based on Deep Learning (DL), especially Convolutional Neural Networks (CNNs), have the ability to learn how to extract and combine relevant representations directly from data. Early DL works on ICBHI, such as LungRN+NL~\cite{lungrn+nl} explored data augmentation to address the data class imbalance , as well as attention mechanism in follow-up work LungAttn~\cite{Li_2021} to improve classification accuracy of respiratory sounds.
In addition to data augmentation, RespireNet~\cite{gairola2020respirenet} uses a model pretrained on ImageNet, with a device specific finetuning strategy. A very recent work~\cite{9729496} instead proposed spectrum correction to scale the frequency responses of the recording devices, as well as a co-tuning strategy to learn the relationship between source and target categories to improve transfer learning. 

Because of class imbalance and different recording settings, there may be hard samples, and training using the cross-entropy loss may be affected by these samples. In this work, we compare cross-entropy training and supervised contrastive training, as done for environmental sound classification task in SoundCLR~\cite{soundclr}, and show that the combination of both frameworks can further improve respiratory sound classification scores. In addition, we learn representations using only data augmentation in the SimCLR~\cite{chen2020simple} framework, and using both data augmentation and available metadata in the SCL framework.

Ideas from SCL have already been tested in the context of ICBHI. Song et al.~\cite{9414385} used a simplified SCL framework, by sampling a first batch of examples, and according to their class labels, a second batch is constructed so that each example has a corresponding positive example. For negative samples, a fixed number of samples with different class labels is chosen from all other samples in the two batches.  However they did not report results on the official split of ICBHI. In this paper, we adapt the original formulation of SCL of constructing pairs, and outperform cross-entropy (CE) training. We test whether a combination of SCL with CE can further boost the classification scores on ICBHI official split. Finally, we show the potential of using available metadata in learning useful representations of respiratory sounds. While using metadata with patient identification with SimCLR has been tested in medical image analysis~\cite{pmlr-v149-vu21a}, we propose here to extend the SCL framework to multiple heads to help disentangle subspaces corresponding to different metadata on both ICBHI and SPRSound.
Our main contributions are summarized as follows : 
\begin{enumerate}
    \item We show that supervised contrastive learning outperforms the cross-entropy training for respiratory sound classification when finetuning a model pretrained on AudioSet.
    \item We combine supervised contrastive learning with cross-entropy and show that it outperforms cross-entropy in correctly classifying anomalies (higher sensitivity).
    \item We show that supervised contrastive using metadata with or without class labels learn useful representations, and propose an extension with multiple pretext tasks for a performance boost.
\end{enumerate}

\section{Methods}
\label{sec:methods}

Let $f$ be a Neural Network (NN) encoder, $g$ a NN classifier, $x_{i}\in X$ the input breathing cycle, and $y_{i}\in Y$ the class label, the cross entropy loss (CE) is calculated as follows :
\begin{equation}
\label{ce}
\mathcal{L}^{CE}  = - \sum_{i} y_{i} log(g(f(A(x_{i}))))
\end{equation}
where $A$ is a stochastic augmentation function and $i\in \{1...N\}$ with $N$ being batch size.
Here, we train ${g~\circ~f}$ to predict the respiratory breathing class for a given respiratory cycle (Fig~\ref{fig:pip}a).

Supervised contrastive learning (SCL) consists of learning a classification task in two steps : first, the feature extractor is trained to pull together in the embedding space samples with the same label, and to push away samples with different label. Second, the linear classifier is trained on the frozen representations learned in the first step (Fig~\ref{fig:pip}b). Formally, the first step consists in adding to the encoder $f$ a shallow NN $h$ called a projector (usually a MLP with one hidden layer) that maps representations to the space where the contrastive loss is applied. In the second step, $h$ is discarded (representations before the non linear projector contains more information~\cite{chen2020simple}), then a classifier $g'$ is trained on the frozen representations (output of $f$) trained in the first step. The supervised contrastive loss (SCL) is calculated as follows :
\begin{equation}
\label{scl}
\mathcal{L}^{SCL} = \sum_{i\in I}\frac{-1}{|P(i)|}\sum_{p\in P(i)}\log{\frac{\text{exp}\left(\boldsymbol{z}_i\boldsymbol{\cdot}\boldsymbol{z}_p/\tau\right)}{\sum\limits_{n\in N(i)}\text{exp}\left(\boldsymbol{z}_i\boldsymbol{\cdot}\boldsymbol{z}_n/\tau\right)}}
\end{equation}

where $i\in I=\{1...2 N\}$ the index of an augmented sample within a training batch.
$\boldsymbol{z}_{i}=h(f(A(\boldsymbol{x}_{i})))\in\mathbb{R}^{D_P}$ where ${D_P}$ is the projector's dimension. ${P(i)={\{p\in I:{{y}}_p={{y}}_i}\}}$ is the set of indices of all positives in the multiviewed batch distinct from $i$ sharing similar label with $i$, and $|P(i)|$ is its cardinality, ${N(i)={\{n\in I:{{y}}_n\neq{{y}}_i}\}}$ is the set of indices of all negatives in the multiviewed batch having dissimilar label with $i$, the $\boldsymbol{\cdot}$ symbol denotes the dot product, and $\tau\in\mathbb{R}^{+*}$ is a scalar temperature parameter.

We extend the SCL framework to solving multiple pretext tasks using multiple heads sharing the same backbone, we call it Multi-Supervised Contrastive (M-SCL)~(Fig~\ref{fig:pip}d), we define its loss $\mathcal{L}^{M-SCL}$ as :
\begin{equation}
\label{mscl}
\mathcal{L}^{M-SCL} = \sum_{i\in \{1...K\}}\lambda_{i} \mathcal{L}_{i}^{SCL}
\end{equation}

where $K$ the number of pretext tasks, $h_{i}$ the projector for the $i^{th}$ task and $\lambda_{i}$ the coefficient for the $i^{th}$ loss $\mathcal{L}_{i}^{SCL}$.

We also consider the hybrid approach, that combines both CE loss and SCL loss (Fig~\ref{fig:pip}c), and minimize the folowing hybrid loss term :

\begin{equation}
\label{hybrid}
    \mathcal{L}^{Hybrid} = \alpha \mathcal{L}^{CE}  + (1 - \alpha) \mathcal{L}^{SCL}
\end{equation}
where $\alpha$ controls the tradeoff between the CE and SCL loss terms.

In both M-SCL and Hybrid settings, the backbone $f$ is shared between heads.

\begin{figure}[!t]
\begin{minipage}[b]{1.0\linewidth}
  \centering
  \centerline{\includegraphics[width=1.\textwidth]{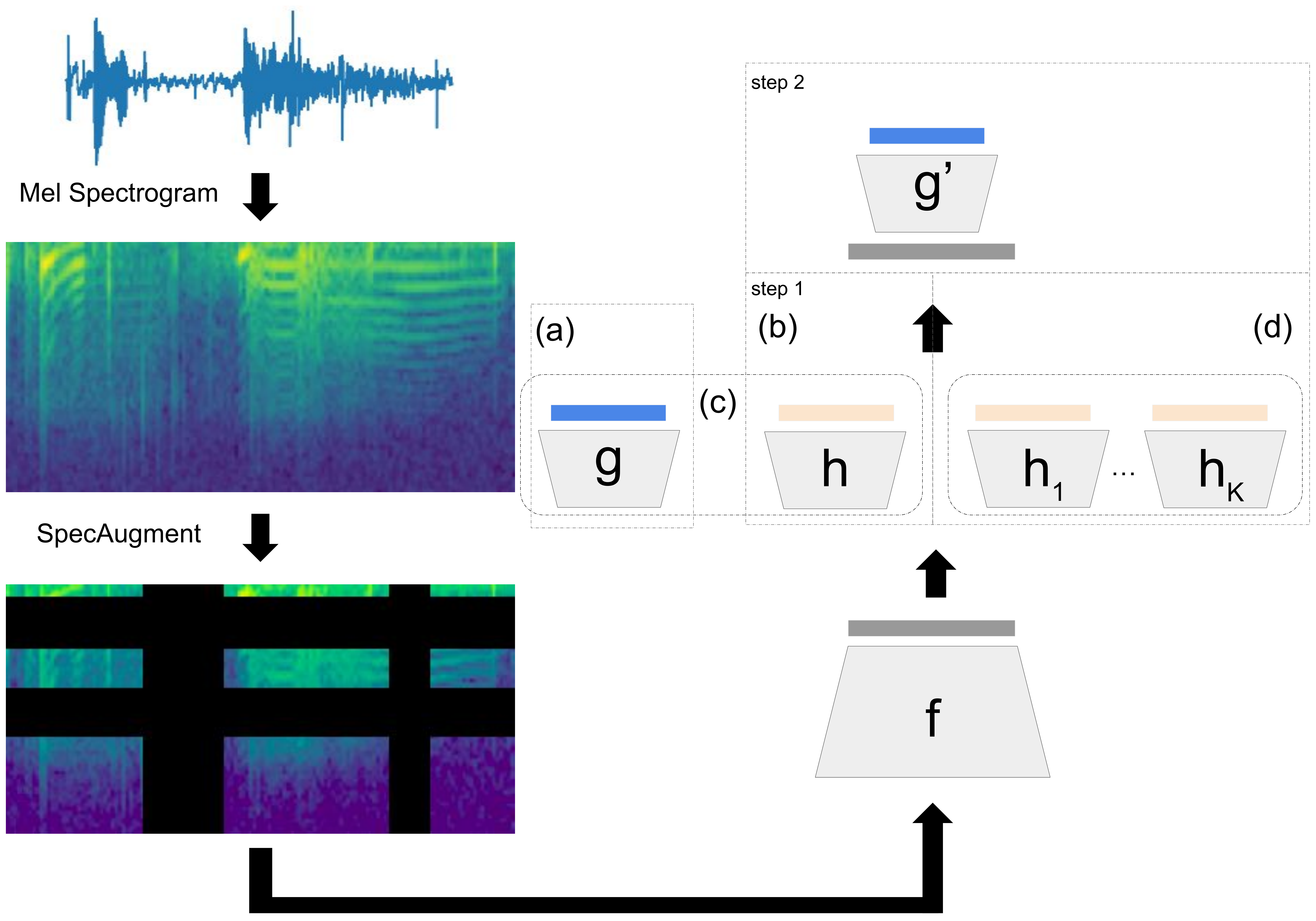}}
%  \vspace{2.0cm}
\end{minipage}
\caption{Overview of the proposed framework : cross-entropy training (a), supervised contrastive learning (b), hybrid training (c), multi-supervised contrastive learning (d).}
\label{fig:pip}
\end{figure}

\section{Experiments}
\label{sec:exps}

\subsection{Datasets}
\label{ssec:dataset}

{\bf ICBHI 2017}~\cite{rocha2019open} is a dataset of the 2017 ICBHI challenge that consists of 5.5 hours of recordings containing 6898 respiratory cycles with a duration ranging from 0.2s to 16.2s (mean cycle duration is 2.7s). 3642 respiratory cycles contain normal breathing, 1864 contain crackles, 886 contain wheezes, and 506 contain both crackles and wheezes, in 920 audio samples from 126 subjects. {\bf SPRSound}~\cite{SPRSound} is an open-source paediatric respiratory sound database consisting of 2,683 records and 9,089 respiratory sound events from 292 participants. Specifically, there are 6887 normal breathing cycles, 53 rhonchi, 865 wheeze, 17 stridor, 66 coarse crackle, 1167 fine crackle, and 34 of both crackle and wheeze events. There are two tests splits : intra-patient and inter-patient. We consider only the inter-patient test split to test for generalization capabilities of learned representations to unseen patients, a more realistic use case in a clinical setting. These are the two largest respiratory sound datasets, containing additional metadata about patients and acquisitions suited to test our approach of learning representations using metadata.

\subsection{Preprocessing and Data Augmentation}
\label{ssec:preprocess_da}

The audio recordings are sampled with a rate that varies from 4 kHz to 44.1 kHz for ICBHI and 8 kHz for SPRSound. We re-sample all recordings to 16 kHz mono as done in previous studies~\cite{9414385,9729496}. We uniformly limit the maximum duration of a respiratory cycle to 8 seconds. We convert the audio signal into the time-frequency representation Mel-spectrogram, with 64 Mel filterbanks, a window size of 1024 over a hop size of 512, with a minimum and a maximum frequency of 50 and 2000 Hz respectively, because wheezes and crackles are in this interval~\cite{10.1007/978-981-10-7419-6_7}. We use the data augmentation method SpecAugment~\cite{park19e_interspeech}, that consists of masking blocks of frequency channels and time steps, followed by time warping, that help the network learn features that are robust to partial loss of frequency and time information, and to deformation in the time direction. (Fig~\ref{fig:pip}). SpecAugment has been tested in self-supervised learning of audio representations in previous works~\cite{fonseca2021unsupervised}.

\subsection{Model}
\label{ssec:model}

\begin{table*}[h]
\caption{Performance analysis on ICBHI.}
\label{Tab:sota}
\begin{center}
\scalebox{0.8}{
\begin{tabular}{|cc|ccccc|}
\hline
\multicolumn{2}{|c|}{Method}                                                  & \multicolumn{1}{c|}{$Sp$}           & \multicolumn{1}{c|}{$Se$}           & \multicolumn{1}{c|}{$Sc$}           & \multicolumn{1}{c|}{\# of Parameters (in M)} & Ext. Dataset \\ \hline%\hline
\multicolumn{2}{|c|}{LungRN+NL\cite{lungrn+nl}}              & \multicolumn{1}{c|}{63.2}           & \multicolumn{1}{c|}{\textbf{41.3}}           & \multicolumn{1}{c|}{52.3}           & \multicolumn{1}{c|}{-}                       & None            \\ %\hline
\multicolumn{2}{|c|}{LungAttn\cite{Li_2021}}              & \multicolumn{1}{c|}{\textbf{71.44}}           & \multicolumn{1}{c|}{36.36}           & \multicolumn{1}{c|}{\textbf{53.9}}           & \multicolumn{1}{c|}{0.7}                       & None            \\ \hline%\hline
\multicolumn{2}{|c|}{Wang et al.\cite{9746941}}              & \multicolumn{1}{c|}{70.4}           & \multicolumn{1}{c|}{40.2}           & \multicolumn{1}{c|}{55.3}           & \multicolumn{1}{c|}{25 (estimated)}          & ImageNet     \\ %\hline
\multicolumn{2}{|c|}{RespireNet\cite{gairola2020respirenet}} & \multicolumn{1}{c|}{72.3}           & \multicolumn{1}{c|}{40.1}           & \multicolumn{1}{c|}{56.2}           & \multicolumn{1}{c|}{21 (estimated)}          & ImageNet     \\ %\hline
\multicolumn{2}{|c|}{ARSC-Net\cite{9669787}}                 & \multicolumn{1}{c|}{67.13}          & \multicolumn{1}{c|}{\textbf{46.38}}          & \multicolumn{1}{c|}{56.76}          & \multicolumn{1}{c|}{-}          & ImageNet     \\ %\hline
\multicolumn{2}{|c|}{Nguyen et al.\cite{9729496} (Vanilla)}            & \multicolumn{1}{c|}{76.33}          & \multicolumn{1}{c|}{37.37}          & \multicolumn{1}{c|}{56.85}          & \multicolumn{1}{c|}{23 (estimated)}          & ImageNet     \\ %\hline
\multicolumn{2}{|c|}{Nguyen et al.\cite{9729496} (StochNorm)}            & \multicolumn{1}{c|}{78.86}          & \multicolumn{1}{c|}{36.40}          & \multicolumn{1}{c|}{57.63}          & \multicolumn{1}{c|}{23 (estimated)}          & ImageNet     \\ %\hline
\multicolumn{2}{|c|}{Nguyen et al.\cite{9729496} (CoTuning)}            & \multicolumn{1}{c|}{\textbf{79.34}}          & \multicolumn{1}{c|}{37.24}          & \multicolumn{1}{c|}{\textbf{58.29}}          & \multicolumn{1}{c|}{23 (estimated)}          & ImageNet     \\ \hline%\hline
\multicolumn{1}{|c|}{Backbone}                            & Method            & \multicolumn{5}{c|}{Our Results (10 runs)}                                                                                                                                              \\ \hline%\hline
\multicolumn{1}{|c|}{\multirow{3}{*}{CNN6}}               & CE                & \multicolumn{1}{c|}{\textbf{76.72$\pm$3.97}} & \multicolumn{1}{c|}{31.12$\pm$3.72} & \multicolumn{1}{c|}{53.92$\pm$0.71} & \multicolumn{1}{c|}{4.3}                     & None            \\ %\cline{2-7} 
\multicolumn{1}{|c|}{}                                    & SCL               & \multicolumn{1}{c|}{76.17$\pm$3.84} & \multicolumn{1}{c|}{27.97$\pm$3.92} & \multicolumn{1}{c|}{52.08$\pm$1.06} & \multicolumn{1}{c|}{4.3}                     & None            \\ %\cline{2-7} 
\multicolumn{1}{|c|}{}                                    & Hybrid            & \multicolumn{1}{c|}{75.35$\pm$5.47} & \multicolumn{1}{c|}{\textbf{33.84$\pm$5.67}} & \multicolumn{1}{c|}{\textbf{54.74$\pm$0.5}}  & \multicolumn{1}{c|}{4.3}                     & None            \\ \hline%\hline
\multicolumn{1}{|c|}{\multirow{3}{*}{CNN6}}               & CE                & \multicolumn{1}{c|}{70.09$\pm$3.08} & \multicolumn{1}{c|}{40.39$\pm$2.97} & \multicolumn{1}{c|}{55.24$\pm$0.43} & \multicolumn{1}{c|}{4.3}                     & AudioSet     \\ %\cline{2-7} 
\multicolumn{1}{|c|}{}                                    & SCL               & \multicolumn{1}{c|}{\textbf{75.95$\pm$2.31}} & \multicolumn{1}{c|}{39.15$\pm$1.89} & \multicolumn{1}{c|}{\textbf{57.55$\pm$0.81}} & \multicolumn{1}{c|}{4.3}                     & AudioSet     \\ %\cline{2-7} 
\multicolumn{1}{|c|}{}                                    & Hybrid            & \multicolumn{1}{c|}{70.47$\pm$2.07} & \multicolumn{1}{c|}{\textbf{43.29$\pm$1.83}} & \multicolumn{1}{c|}{56.89$\pm$0.55} & \multicolumn{1}{c|}{4.3}                     & AudioSet     \\
 \hline
 \multicolumn{7}{c}{\scriptsize *We highlights in bold our best scores as well as best scores from the litterature both from scratch and from pretraining.} \\
\end{tabular}
}
\end{center}
\end{table*}

We use models introduced for the dataset AudioSet~\cite{Kong2020PANNsLP}, a large dataset that contains 2 million sounds including respiratory classes. We compare results with and without pretraining on Audioset. 
We report results with the CNN6 model, that consists in 4 blocks, each containing a 2D convolution with a kernel size of 5, a batch normalization and an average pooling with a kernel size of 2. CNN6  contains 4.3 million parameters, compared to other works on ICBHI that use ResNet34, ResNet50, and ResNet101 containing approximately 21, 25, and 44 million parameters, respectively\cite{gairola2020respirenet,9729496}.

\subsection{Evaluation Metrics}
\label{ssec:eval}

We adopt the same evaluation metrics as the official ICBHI 2017 and SPRS challenges :
\begin{equation}
\label{se}
\begin{split}
    Se=\frac{TP}{TP+FN};\,Sp=\frac{TN}{TN+FP}; \\ Sc=\frac{Se+Sp}{2};\, HS=2*\frac{Se*Sp}{Se+Sp};
\end{split}
\end{equation}
where $TP$, $TN$, $FP$, and $FN$ stand for the numbers of true positives, true negatives, false positives and false negatives, respectively.

\subsection{Experimental Setting}
\label{ssec:exp}

We experiment with different setups : CE, SCL, Hybrid, with and without AudioSet pretraining, and M-SCL with 2 pretext tasks with Audioset pretraining, the first pretext task uses respiratory classes, and the second uses synthetic metadata classes (4) : age group (Old/Young for ICBHI and Kid/Baby for SPRSound) and sex (M/F) as these factors can be correlated with presence of respiratory anomalies~\cite{clhls}. We modified the original CNN6 model as follows: we remove dropout, keep all convolutional and pooling layers, then add a simple linear layer for the classifier in the CE setting and the first head of the hybrid setting, and add a MLP with one hidden layer with 128 neurons for the SCL, M-SCL, and the second head of the hybrid setting. We train all models on an Nvidia 3090 GPU for 400 epochs on ICBHI and for 200 epochs on SPRSound (the model converges on it faster) using Adam optimizer with a momentum of 0.9, a batch size of 128 with initial learning rate of $10^{-4}$ for pretrained models and a learning rate of $10^{-3}$ for models trained from scratch, and a weight decay of $10^{-4}$. We use cosine annealing as a learning rate schedule without warm restarts for all the experiments, except the second phase of supervised contrastive where we train a linear classifier on frozen representations using a learning rate of 0.1 without scheduling.
We performed a grid search to find the best parameters for SpecAugment on ICBHI. The best and most stable results were obtained without time warping, when masking two blocks of twenty consecutive mel frequency channels, and two blocks of forty consecutive time steps. We also performed a grid search to find the best value of $\tau=0.06$ in SCL as well as $\alpha=0.5$ for hybrid training, and $\lambda_{1}=0.25$ and $\lambda_{2}=0.75$ for M-SCL. We use the same values of hyperparameters for SPRSound. We report results on the ICBHI official challenge split and the SPRSound inter-patient set (as the intra-patient set doesn't reflect generalization capabilities). In order to be comparable with previous approaches, and in the absence of validation split, results correspond to the epoch with the best $Sc$ on the test split as done by previous works~\cite{gairola2020respirenet}. For stability and robustness, we report results over ten identical runs.

\subsection{Results}
\label{ssec:results}
Table~\ref{Tab:sota} is composed of 2 panels : the upper panel shows state-of-the-art performance on the ICBHI official split, and the bottom panel shows our results across 10 identical runs using cross-entropy training, supervised contrastive training with class labels only, and hybrid training for CNN6 both from scratch and with AudioSet pretraining. Our results show that supervised contrastive learning or hybrid learning surpass cross-entropy training when training from scratch or by finetuning from AudioSet. A fine-tuned CNN6 reaches a score of 57.55 in the supervised contrastive setting outperforming all previous work except the setting of the recent work of Nguyen et al.~\cite{9729496}, in which they introduced co-tuning technique and obtains a score of 58.29, while our approach compares well with their vanilla finetuning and StochNorm scores. In terms of memory requirement, our approach has up to five times less parameters than previously proposed approaches with networks finetuned from Imagenet. Finally, it is worth noting that our approach with CNN6 trained from scratch with the hybrid loss achieves the new state of the art on ICBHI when not considering pretraining on an external dataset.

\begin{table}[]
\caption{Results on IBCHI using metadata}
\label{Tab:tab2}
\resizebox{\columnwidth}{!}{%
\begin{tabular}{lccc}
\hline
Method                                                & $Sp$          & $Se$          & $Sc$          \\ \hline
\multicolumn{4}{l}{\textit{without respiratory classes}}                                                      \\ \hline
SimCLR                                            & 59.75$\pm$5.81 & 36.93$\pm$5.71 & 48.34$\pm$0.87 \\
SCL w/ Sex+Age                                                   & 71.25$\pm$2.79) & 39.32$\pm$3.35 & 55.29$\pm$0.80 \\
 \hline
\multicolumn{4}{l}{\textit{with respiratory classes}}                                                         \\ \hline
SCL w/ Age+Class                                                    & 70.84$\pm$3.19 & \textbf{40.47$\pm$3.84} & 55.65$\pm$1.27 \\
SCL w/ Sex+Class                                                    & 71.00$\pm$4.52 & 39.72$\pm$3.13 & 55.36$\pm$1.00 \\
SCL w/ Sex+Age+Class                                                   & 71.56$\pm$4.03 & 40.06$\pm$3.09 & 55.81$\pm$0.88 \\
M-SCL w/ Sex+Age \& Class  & \textbf{76.93$\pm$2.99} & 39.15$\pm$2.84 & \textbf{58.04$\pm$0.94} \\
 \hline
\end{tabular}
}
\end{table}

For learning representations using metadata, we consider the pretext task of pulling together respiratory cycles sharing the same demographics (sex, and/or age group) while pushing away dissimilar ones in the latent space. We also test a combination of metadata with class labels, using either a single head with a label combining target classes and metadata using SCL or multiple heads (M-SCL). In a second step, we train a linear classifier on the frozen representation using class labels. Table~\ref{Tab:tab2} shows the potential of learning representations of data with or without the use of class labels; the first panel shows that using sex and age group with SCL learns representations that obtain a score of 55.29$\pm$0.80, outperforming the SSL baseline SimCLR (48.34$\pm$0.87), and doing as well as cross-entropy training with class labels (Table~\ref{Tab:sota}); in the second panel, we can see that M-SCL (58.04$\pm$0.94) outperforms SCL (55.81$\pm$0.88), we note here that both methods use metadata and class labels : M-SCL uses a first head head for class labels and a second head for metadata, while SCL uses one head for the label obtained by combining metadata and classlabels. SCL with label and metadata performs worse than SCL with class labels from Table~\ref{Tab:sota}, we hypothesize that it's because for some synthetic classes only few samples are available (e.g. only 3 respiratory cycles with crackles and wheezes come from young female), therefore it's hard to learn discriminative feature for those classes. However, M-SCL outperforms our best result in Table~\ref{Tab:sota} (SCL : 57.55$\pm$0.81), showing the performance boost obtained from leveraging metadata.

\begin{table}[]
\caption{Results on SPRSound}
\label{Tab:sprs}
\resizebox{\columnwidth}{!}{%
\begin{tabular}{ccccc}
\hline
Method   & $Se$             & $Sp$             & $Sc$             & $HS$             \\ \hline
CE       & 76.89$\pm$0.80 & 92.35$\pm$1.10 & 84.62$\pm$0.29 & 83.90$\pm$0.25 \\
SCL      & 80.69$\pm$1.62 & 90.49$\pm$1.27 & 85.59$\pm$0.48 & 85.29$\pm$0.56 \\
M-SCL    & 82.24$\pm$2.24 & 88.62$\pm$1.48 & 85.43$\pm$0.79 & 85.27$\pm$0.88 \\
Baseline & 51.93          & 77.88          & 64.90          & 62.31          \\ \hline
\end{tabular}
}
%{\scriptsize *We highlights in bold the two highest scores.}
\end{table}

Table~\ref{Tab:sprs} shows the performance of CNN6 (pretrained on AudioSet) when finetuned on SPRSound using CE, SCL or M-SCL, compared to the baseline of Naive Bayes Classifier trained on the mel-frequency cepstral coefficient (MFCC). Our experiments show that SCL and M-SCL outperform CE in both harmonic score ($HS$) and mean score ($Sc$), only the specificity ($Sp$) is higher for CE, we assume this is due to overfitting the normal breathing cycle because of the imbalance in the dataset. On the contrary, our contrastive approach learns to better separate anomalies in the latent space as the sensitivity ($Se$) is higher than of CE, especially for M-SCL when leveraging metadata information for learning representations. We did not compare ours results on SPRS to other works other than the baseline, because the few published works report their results by combining both inter and intra sets.

On both datasets, contrastive learning has led to higher score than cross entropy training with high sensitivity. Contrastive learning learns to cluster similar breathings in the latent space while pushing apart dissimilar ones, a desirable property for classification. Metadata provided additional information to be taken into account when learning the representations of breathings. We experimented with sex and age (4 groups in total), and they either boosted sensitivity (on SPRSound) or overall score (on ICBHI). Our approach is distinct from previous work that have used contrastive approaches for respiratory sounds; the work of~\cite{clhls} trained a system on ICBHI to diagnose patients, which is an easier task. \cite{9414385} have used SCL on ICBHI with different sampling of negatives examples, as well as a different cross-validation split, making it difficult to compare. Overall, we have proposed a supervised contrastive approach that exploits metadata in a simple, effective and reproducible way on two datasets. 

\section{Conclusion}
\label{sec:end}

We show in this work the potential of supervised contrastive learning for an imbalanced and noisy setting, outperforming cross-entropy using experiments on respiratory sound classification. We also show that using metadata to combine multiple supervised contrastive tasks for learning useful representations obtain state-of-the-art results.
In future work, we will attempt at building upon the multi-head framework using several pretext tasks such as exploiting spatial or temporal metadata associated with recordings, and investigate data augmentation techniques that better address the variability in the low data regime. Ultimately, such approaches could be adapted to generalize to unseen auditory tasks such as detection or localization, and deal with larger domain shifts.

\section{Acknowledgment}
This work was co-funded by the AI@IMT program of the ANR (French National Research Agency) and the company OSO-AI. We would like to thank our colleagues in the BRAIn Team of the Mathematical and Electrical Engineering Department of Institut Mines-Télécom Atlantique for their insights and feedback.

%\pagebreak
% -------------------------------------------------------------------------
% Either list references using the bibliography style file IEEEtran.bst
\bibliographystyle{IEEEtran}
\bibliography{refs23}

\begin{thebibliography}{10}
\providecommand{\url}[1]{#1}
\def\UrlFont{\rmfamily}
\providecommand{\newblock}{\relax}
\providecommand{\bibinfo}[2]{#2}
\providecommand\BIBentrySTDinterwordspacing{\spaceskip=0pt\relax}
\providecommand\BIBentryALTinterwordstretchfactor{4}
\providecommand\BIBentryALTinterwordspacing{\spaceskip=\fontdimen2\font plus
\BIBentryALTinterwordstretchfactor\fontdimen3\font minus
  \fontdimen4\font\relax}
\providecommand\BIBforeignlanguage[2]{{%
\expandafter\ifx\csname l@#1\endcsname\relax
\typeout{** WARNING: IEEEtran.bst: No hyphenation pattern has been}%
\typeout{** loaded for the language `#1'. Using the pattern for}%
\typeout{** the default language instead.}%
\else
\language=\csname l@#1\endcsname
\fi
#2}}

\bibitem{chen2020simple}
T.~Chen, S.~Kornblith, M.~Norouzi, and G.~Hinton, ``A simple framework for
  contrastive learning of visual representations,'' in \emph{International
  conference on machine learning}, 2020.

\bibitem{fonseca2021unsupervised}
E.~Fonseca, D.~Ortego, K.~McGuinness, N.~E. O'Connor, and X.~Serra,
  ``Unsupervised contrastive learning of sound event representations,'' in
  \emph{2021 IEEE International Conference on Acoustics, Speech and Signal
  Processing (ICASSP)}, 2021.

\bibitem{fonseca2021self}
E.~Fonseca, A.~Jansen, D.~P. Ellis, S.~Wisdom, M.~Tagliasacchi, J.~R. Hershey,
  M.~Plakal, S.~Hershey, R.~C. Moore, and X.~Serra, ``Self-supervised learning
  from automatically separated sound scenes,'' in \emph{2021 IEEE Workshop on
  Applications of Signal Processing to Audio and Acoustics (WASPAA)}, 2021.

\bibitem{cola}
A.~Saeed, D.~Grangier, and N.~Zeghidour, ``Contrastive learning of
  general-purpose audio representations,'' in \emph{ICASSP 2021 - 2021 IEEE
  International Conference on Acoustics, Speech and Signal Processing
  (ICASSP)}, 2021, pp. 3875--3879.

\bibitem{khosla2020supervised}
P.~Khosla, P.~Teterwak, C.~Wang, A.~Sarna, Y.~Tian, P.~Isola, A.~Maschinot,
  C.~Liu, and D.~Krishnan, ``Supervised contrastive learning,'' \emph{Advances
  in Neural Information Processing Systems}, 2020.

\bibitem{soundclr}
A.~Nasiri and J.~Hu, ``Soundclr: Contrastive learning of representations for
  improved environmental sound classification,'' 2021, \text{arXiv:2103.01929}.

\bibitem{rocha2019open}
B.~M. Rocha, D.~Filos, L.~Mendes, G.~Serbes, S.~Ulukaya, Y.~P. Kahya,
  N.~Jakovljevic, T.~L. Turukalo, I.~M. Vogiatzis, E.~Perantoni, \emph{et~al.},
  ``An open access database for the evaluation of respiratory sound
  classification algorithms,'' \emph{Physiological measurement}, 2019.

\bibitem{SPRSound}
Q.~Zhang, J.~Zhang, J.~Yuan, H.~Huang, Y.~Zhang, B.~Zhang, G.~Lv, S.~Lin,
  N.~Wang, X.~Liu, M.~Tang, Y.~Wang, H.~Ma, L.~Liu, S.~Yuan, H.~Zhou, J.~Zhao,
  Y.~Li, Y.~Yin, L.~Zhao, G.~Wang, and Y.~Lian, ``Sprsound: Open-source sjtu
  paediatric respiratory sound database,'' \emph{IEEE Transactions on
  Biomedical Circuits and Systems}, vol.~16, no.~5, pp. 867--881, 2022.

\bibitem{lungrn+nl}
Y.~Ma, X.~Xu, and Y.~Li, ``Lungrn+nl: An improved adventitious lung sound
  classification using non-local block resnet neural network with mixup data
  augmentation,'' in \emph{Proc. Interspeech 2020}, 2020.

\bibitem{Li_2021}
\BIBentryALTinterwordspacing
J.~Li, J.~Yuan, H.~Wang, S.~Liu, Q.~Guo, Y.~Ma, Y.~Li, L.~Zhao, and G.~Wang,
  ``Lungattn: advanced lung sound classification using attention mechanism with
  dual tqwt and triple stft spectrogram,'' \emph{Physiological Measurement},
  2021. [Online]. Available: \url{https://dx.doi.org/10.1088/1361-6579/ac27b9}
\BIBentrySTDinterwordspacing

\bibitem{gairola2020respirenet}
S.~Gairola, F.~Tom, N.~Kwatra, and M.~Jain, ``Respirenet: A deep neural network
  for accurately detecting abnormal lung sounds in limited data setting,''
  2020.

\bibitem{9729496}
T.~Nguyen and F.~Pernkopf, ``Lung sound classification using co-tuning and
  stochastic normalization,'' \emph{IEEE Transactions on Biomedical
  Engineering}, 2022.

\bibitem{9414385}
W.~Song, J.~Han, and H.~Song, ``Contrastive embeddind learning method for
  respiratory sound classification,'' in \emph{ICASSP 2021 - 2021 IEEE
  International Conference on Acoustics, Speech and Signal Processing
  (ICASSP)}, 2021.

\bibitem{pmlr-v149-vu21a}
\BIBentryALTinterwordspacing
Y.~N.~T. Vu, R.~Wang, N.~Balachandar, C.~Liu, A.~Y. Ng, and P.~Rajpurkar,
  ``Medaug: Contrastive learning leveraging patient metadata improves
  representations for chest x-ray interpretation,'' in \emph{Proceedings of the
  6th Machine Learning for Healthcare Conference}, 2021. [Online]. Available:
  \url{https://proceedings.mlr.press/v149/vu21a.html}
\BIBentrySTDinterwordspacing

\bibitem{10.1007/978-981-10-7419-6_7}
N.~Jakovljevi{\'{c}} and T.~Lon{\v{c}}ar-Turukalo, ``Hidden markov model based
  respiratory sound classification,'' in \emph{Precision Medicine Powered by
  pHealth and Connected Health}, 2018.

\bibitem{park19e_interspeech}
D.~S. Park, W.~Chan, Y.~Zhang, C.-C. Chiu, B.~Zoph, E.~D. Cubuk, and Q.~V. Le,
  ``{SpecAugment: A Simple Data Augmentation Method for Automatic Speech
  Recognition},'' in \emph{Proc. Interspeech 2019}, 2019.

\bibitem{9746941}
Z.~Wang and Z.~Wang, ``A domain transfer based data augmentation method for
  automated respiratory classification,'' in \emph{ICASSP 2022 - 2022 IEEE
  International Conference on Acoustics, Speech and Signal Processing
  (ICASSP)}, 2022.

\bibitem{9669787}
L.~Xu, J.~Cheng, J.~Liu, H.~Kuang, F.~Wu, and J.~Wang, ``Arsc-net: Adventitious
  respiratory sound classification network using parallel paths with
  channel-spatial attention,'' in \emph{2021 IEEE International Conference on
  Bioinformatics and Biomedicine (BIBM)}, 2021.

\bibitem{Kong2020PANNsLP}
Q.~Kong, Y.~Cao, T.~Iqbal, Y.~Wang, W.~Wang, and M.~D. Plumbley, ``Panns:
  Large-scale pretrained audio neural networks for audio pattern recognition,''
  \emph{IEEE/ACM Transactions on Audio, Speech, and Language Processing}, 2020.

\bibitem{clhls}
\BIBentryALTinterwordspacing
P.~N. Soni, S.~Shi, P.~R. Sriram, A.~Y. Ng, and P.~Rajpurkar, ``Contrastive
  learning of heart and lung sounds for label-efficient diagnosis,'' 2021.
  [Online]. Available:
  \url{https://www.sciencedirect.com/science/article/pii/S2666389921002671}
\BIBentrySTDinterwordspacing

\end{thebibliography}
%
% or list them by yourself
% \begin{thebibliography}{9}
%
% \bibitem{waspaaweb}
%   \url{http://www.waspaa.com}.
%
% \bibitem{IEEEPDFSpec}
%   {PDF} specification for {IEEE} {X}plore$^{\textregistered}$,
%   \url{http://www.ieee.org/portal/cms_docs/pubs/confstandards/pdfs/IEEE-PDF-SpecV401.pdf}.
%
% \bibitem{PDFOpenSourceTools}
%   Creating high resolution {PDF} files for book production with
%   open source tools,
%   \url{http://www.grassbook.org/neteler/highres_pdf.html}.
%
% \bibitem{eWilliams1999}
% E. Williams, \emph{Fourier Acoustics: Sound Radiation and Nearfield Acoustic
%   Holography}. London, UK: Academic Press, 1999.
%
% \bibitem{ieeecopyright}
%   \url{http://www.ieee.org/web/publications/rights/copyrightmain.html}.
%
% \bibitem{cJones2003}
% C. Jones, A. Smith, and E. Roberts, ``A sample paper in conference
%   proceedings,'' in \emph{Proc. IEEE ICASSP}, vol. II, 2003, pp. 803--806.
%
% \bibitem{aSmith2000}
% A. Smith, C. Jones, and E. Roberts, ``A sample paper in journals,''
%   \emph{IEEE Trans. Signal Process.}, vol. 62, pp. 291--294, Jan. 2000.
%
% \end{thebibliography}

\end{sloppy}
\end{document}